\documentclass[%
 reprint,
superscriptaddress,
 amsmath,amssymb,
 aps,
prl,
]{revtex4-2}

\usepackage{graphicx}
\usepackage{dcolumn}
\usepackage{bm}
\usepackage{float}

\begin{document}

\preprint{APS/123-QED}

\title{Soft phonons and ultralow lattice thermal conductivity in the Dirac semimetal $\textrm{Cd}_3\textrm{As}_2$}

\author{Shengying Yue}
 \affiliation{Department of Mechanical Engineering, University of California, Santa Barbara, CA 93106, USA}
 
\author{Hamid T. Chorsi}%
\affiliation{Department of Electrical and Computer Engineering, University of California, Santa Barbara, CA 93106, USA
}%

\author{Manik Goyal}%
\affiliation{Materials Department, University of California, Santa Barbara, CA 93106, USA
}%

\author{Timo Schumann}%
\affiliation{Materials Department, University of California, Santa Barbara, CA 93106, USA
}%

\author{Runqing Yang}
 \affiliation{Department of Mechanical Engineering, University of California, Santa Barbara, CA 93106, USA}

\author{Tashi Xu}
 \affiliation{Department of Mechanical Engineering, University of California, Santa Barbara, CA 93106, USA}

\author{Bowen Deng}
 \affiliation{Department of Mechanical Engineering, University of California, Santa Barbara, CA 93106, USA}
 
\author{Susanne Stemmer}%
\affiliation{Materials Department, University of California, Santa Barbara, CA 93106, USA
}%

\author{Jon A. Schuller}%
\affiliation{Department of Electrical and Computer Engineering, University of California, Santa Barbara, CA 93106, USA
}%

\author{Bolin Liao}
\email{bliao@ucsb.edu} \affiliation{Department of Mechanical Engineering, University of California, Santa Barbara, CA 93106, USA}

\date{\today}

\begin{abstract}
Recently, $\textrm{Cd}_3\textrm{As}_2$ has attracted intensive research interest as an archetypical Dirac semimetal, hosting three dimensional linear-dispersive electronic bands near the Fermi level. Previous studies have shown that single-crystalline $\textrm{Cd}_3\textrm{As}_2$ has an anomalously low lattice thermal conductivity, ranging from 0.3 W/mK to 0.7 W/mK at 300 K, which has been attributed to point defects. In this work, we combine first-principles lattice dynamics calculations and temperature-dependent high-resolution Raman spectroscopy of high-quality single-crystal thin films grown by molecular beam epitaxy to reveal the existence of a group of soft optical phonon modes at the Brillouin zone center of $\textrm{Cd}_3\textrm{As}_2$. These soft phonon modes significantly increase the scattering phase space of heat-carrying acoustic phonons and are the origin of the low lattice thermal conductivity of $\textrm{Cd}_3\textrm{As}_2$. Furthermore, we show that the interplay between the phonon-phonon Umklapp scattering rates and the soft optical phonon frequency explains the unusual non-monotonic temperature dependence of the lattice thermal conductivity of $\textrm{Cd}_3\textrm{As}_2$. Our results further suggest that the soft phonon modes are potentially induced by a Kohn anomaly associated with the Dirac nodes, in analogy to similar, nonetheless weaker, effects in graphene and Weyl semimetals.  

\end{abstract}

\maketitle

Cadmium arsenide ($\textrm{Cd}_3\textrm{As}_2$) is a well-studied electronic material due to its semimetalicity and ultrahigh charge carrier mobility\cite{rosenberg1959cd3as2,schumann2016molecular}. The research interest in $\textrm{Cd}_3\textrm{As}_2$ has been revived recently due to both theoretical prediction\cite{wang2013three} and experimental observations\cite{borisenko2014experimental,liu2014stable,jeon2014landau,neupane2014observation} that $\textrm{Cd}_3\textrm{As}_2$ hosts three-dimensional linear Dirac bands\cite{crassee20183d}. Namely, the electronic structure of $\textrm{Cd}_3\textrm{As}_2$ is a three-dimensional analog to that of graphene. Extensive optical and electrical transport measurements have been conducted on $\textrm{Cd}_3\textrm{As}_2$ to probe the Dirac physics of its electronic states\cite{liang2015ultrahigh,liang2017anomalous,schumann2018observation}. In comparison, less studied aspects of $\textrm{Cd}_3\textrm{As}_2$ are its lattice dynamics and heat transport properties, which are essential for thermal management of actual electronic devices utilizing this material. Furthermore, combining the recently predicted large thermopower of topological semimetals\cite{skinner2018large} and the ultrahigh charge carrier mobility, $\textrm{Cd}_3\textrm{As}_2$ is also a promising candidate for thermoelectric energy conversion applications\cite{wang2018magnetic,hosseini2016large}, where a low thermal conductivity is required for high efficiency. Therefore, a thorough understanding of the lattice dynamics and thermal transport properties of $\textrm{Cd}_3\textrm{As}_2$ is of paramount importance for its practical applications. Moreover, how the lattice degree of freedom couples to topological properties of the electronic structure is of fundamental interest and remains largely unexplored\cite{song2016detecting,rinkel2017signatures,cheng2019large}.

The thermal conductivity of crystalline $\textrm{Cd}_3\textrm{As}_2$ was first measured in the 1960s\cite{spitzer1966anomalous,armitage1969thermal}. Subtracting the electronic contribution from the total thermal conductivity using the Wiedemann-Franz law, researchers found that $\textrm{Cd}_3\textrm{As}_2$ possessed an anomalously low lattice thermal conductivity (0.3 W/mK at 300 K reported by Spitzer et al.\cite{spitzer1966anomalous}). This value is lower than that of amorphous glass and many polymers and in sharp contrast to other crystalline materials with similar atomic mass and crystal structure. A recent study used a strong magnetic field up to 14 Tesla to freeze out electron transport and measured a lattice thermal conductivity of around 0.7 W/mK at 300 K\cite{wang2018magnetic}. More surprisingly, they found the lattice thermal conductivity of $\textrm{Cd}_3\textrm{As}_2$ increases with temperature above roughly 300 K, which runs counter to the familiar $1/T$ dependence of lattice thermal conductivity in crystalline materials due to phonon-phonon Umklapp scatterings. So far, the origin of the anomalously low lattice thermal conductivity of $\textrm{Cd}_3\textrm{As}_2$ and its unusual temperature dependence remains unclear. Recent measurements of another Dirac semimetal, $\textrm{Zr}\textrm{Te}_5$, also revealed highly anisotropic and very low thermal conductivity\cite{zhu2018record}.

In this work, we combine first-principles lattice dynamics calculations and temperature-dependent high-resolution Raman measurement of high-quality $\textrm{Cd}_3\textrm{As}_2$ thin films grown by molecular beam epitaxy (MBE) to reveal the existence of a group of soft optical phonons in $\textrm{Cd}_3\textrm{As}_2$ with anomalously low frequencies at the Brillouin zone center. These soft optical phonons largely increase the scattering phase space of heat-carrying acoustic phonons and are responsible for the reduction of the lattice thermal conductivity. More interestingly, we find that the optical phonon frequency is highly sensitive to the smearing parameter of the electronic Fermi-Dirac distribution in our calculation. This is potentially a signature of Kohn anomaly associated with the Dirac nodes, as has been found in graphene\cite{piscanec2004kohn,lazzeri2006nonadiabatic} and Weyl semimetal tantalum phosphide\cite{nguyen2019discovery}. As a result, the frequency of the soft optical phonon increases with temperature, which reduces phonon-phonon scattering and leads to an increasing lattice thermal conductivity. Our investigation quantitatively reproduces the low lattice thermal conductivity of $\textrm{Cd}_3\textrm{As}_2$ and qualitatively explains its anomalous temperature dependence.
\begin{figure}[htb]
\centering
\graphicspath{{./}}
\includegraphics[width=\columnwidth]{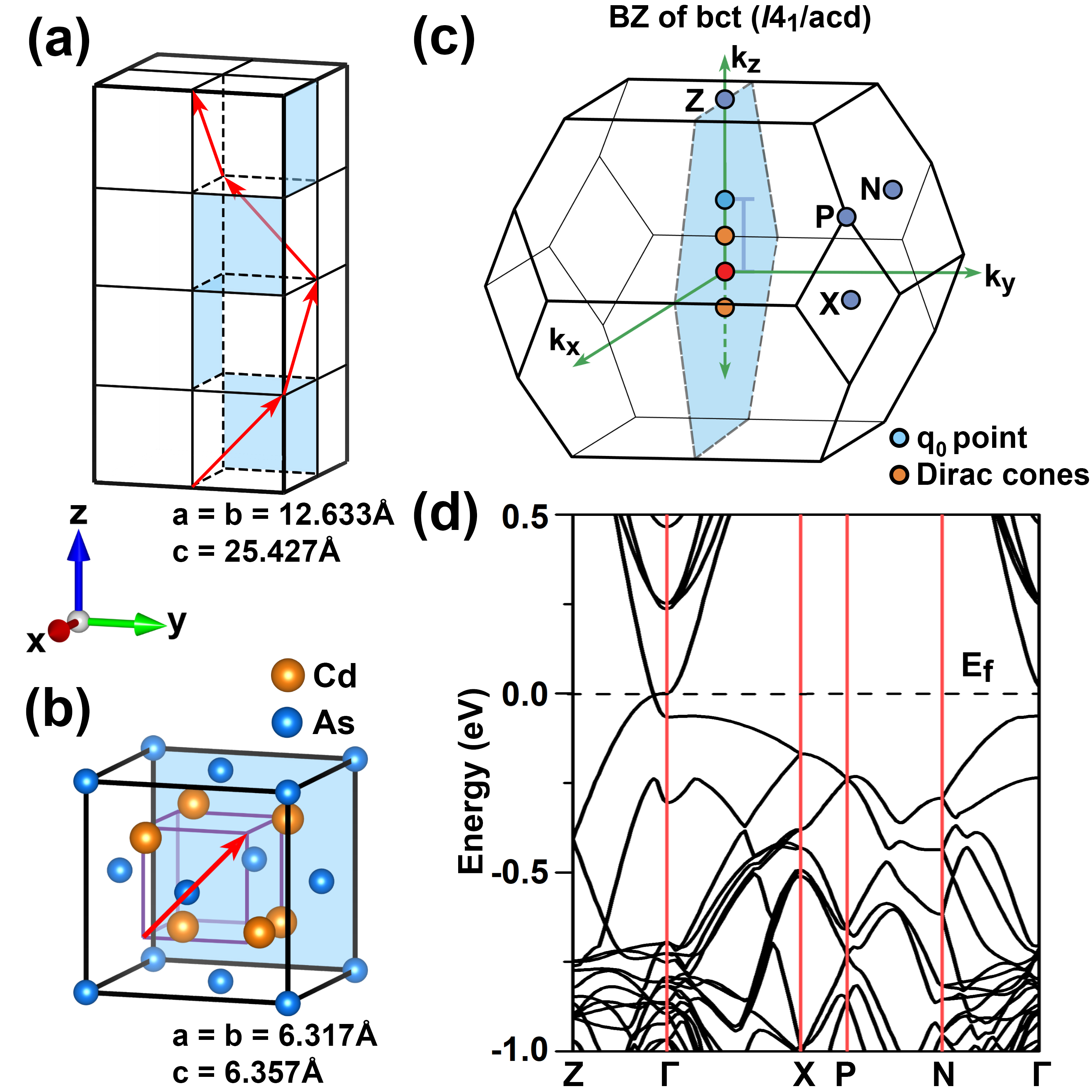}
\caption{Crystal structure and calculated electronic bands of $\textrm{Cd}_3\textrm{As}_2$. (a) Schematic of the $\textrm{Cd}_3\textrm{As}_2$ conventional cell. (b) Structure of the 10-atom building block for the full cell shown in (a). (c) The first Brillouin zone of $\textrm{Cd}_3\textrm{As}_2$, where the location of the Dirac nodes and the position of the Kohn anomaly $\mathbf{q}_0$ are marked. (d) Calculated electronic band structure of $\textrm{Cd}_3\textrm{As}_2$. }
\label{fig1}
\end{figure}

We used the Vienna ab-initio simulation package (VASP)\citep{vasp01,vasp02} to calculate the electronic structure using density functional theory (DFT). The projector augmented wave (PAW)\citep{PAW01,PAW02} method was adopted with the Perdew-Burke-Ernzerhof (PBE)\citep{PBE} exchange correlation functional. The spin-orbit coupling was included in the calculation as implemented in VASP\citep{steiner2016calculation}. The cutoff energy for the plane wave expansion was set to 500 eV. The Monkhorst-Pack k-mesh $\rm 6\times 6 \times 6$ including $\Gamma$ point was used to sample the whole Brillouin zone (BZ). The low-temperature phase of $\textrm{Cd}_3\textrm{As}_2$ below 748 K is body-centered tetragonal (bct) with the space group $\textrm{I}4_1/\textrm{acd}$ and has a conventional cell consisting of 160 atoms\cite{ali2014crystal} (the corresponding primitive cell includes 80 atoms), as shown in Fig. \ref{fig1}(a). The conventional cell can be constructed as a $2\times 2\times 4$ supercell of 10-atom substructures. Each substructure has an antifluorite-derived structure with arsenic atoms at the face-center locations and cadmium atoms that form a cube with two vacancies located diagonally on one surface of the cube\cite{ali2014crystal}, as illustrated in Fig. \ref{fig1}(b). The arrangement of the cadmium vacancies in the substructures alternates in a fashion depicted in Fig. \ref{fig1}(a), where the red arrows mark the position of the cadmium vacancies. Before the electronic band structure calculations, the lattice structure was fully optimized with the Hellmann-Feynman force tolerance $\rm 0.001~ eV/\AA$. The lattice constants of the relaxed conventional cell are given in Fig. \ref{fig1}(a). The calculated electronic structure of $\textrm{Cd}_3\textrm{As}_2$ using the 80-atom primitive cell is shown in Fig. \ref{fig1}(d), where the Dirac cones are located along the $\Gamma$-$Z$ direction at $(0,0,\pm 0.043\  \textrm{\AA}^{-1}$), in agreement with previous reports\cite{ali2014crystal,conte2017electronic,crassee20183d}. The BZ shape, high-symmetry points and the locations of the Dirac cones are shown in Fig. \ref{fig1}(c).

\begin{figure*}[htb]
\centering
\graphicspath{{./}}
 \includegraphics[width=0.8\textwidth]{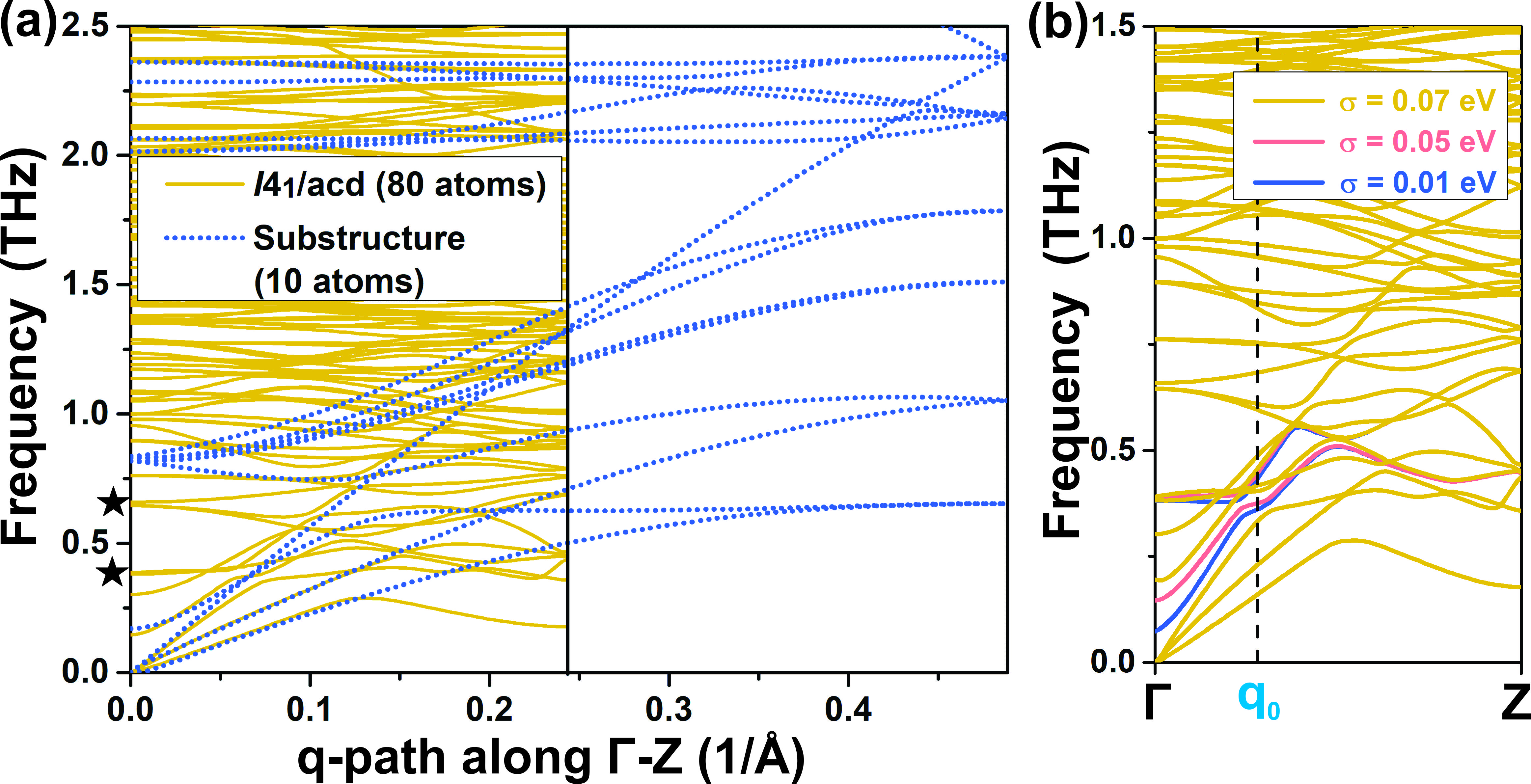}
\caption{(a) Calculated phonon dispersion relations of $\textrm{Cd}_3\textrm{As}_2$, using both the full 80-atom primitive cell and the 10-atom substructure. The stars mark the modes observed in the Raman measurement. (b) The phonon dispersion calculated using the full 80-atom primitive cell showing the dependence of the LOB frequency on the electronic smearing parameter $\sigma$. $\mathbf{q}_0$ marks the position of potential Kohn anomaly in addition to $\Gamma$.}
\label{fig2}
\end{figure*}

Next, we used first-principles calculation to examine the phonon dispersion relations of $\textrm{Cd}_3\textrm{As}_2$, which, to the best of our knowledge, have not been reported before. For this calculation, we obtained the harmonic interatomic force constants (IFCs) by employing the finite displacement method based on DFT force calculations\cite{esfarjani2011heat} implemented in VASP and then diagonalized the dynamical matrices to generate the phonon dispersion relations using PHONOPY\citep{phonopy}. To understand the effect of the short-range arrangement of cadmium vacancies, we calculated and compared the phonon dispersion relations of both the 80-atom bct primitive cell and a crystal constructed by the 10-atom substructure shown in Fig. \ref{fig1}(b) as its primitive cell. The calculated phonon dispersions along the $\Gamma$-Z direction are presented in Fig. \ref{fig2}. The phonon dispersions along other high-symmetry lines in the BZ are given in the Supplementary Information. We notice that the phonon dispersions calculated using the primitive cell and the substructure agree with each other very well near the zone center considering zone folding. This can be explained by the fact that the long-wavelength phonons near the zone center should be less sensitive to the short-range arrangement of the cadmium vacancies. A prominent feature of the phonon dispersions is the presence of a group of soft optical phonon modes at the zone center below 1 THz. In particular, the frequency of the lowest optical branch (LOB) is highly sensitive to the smearing parameter $\sigma$ of the electronic Fermi-Dirac distribution used in the DFT calculation. $\sigma$ (in units of eV) represents the broadening of the electronic occupation around the Fermi level and a fictitious electronic temperature $T_{\textrm{el}}=\sigma / k_B$ can be defined, where $k_B$ is the Boltzmann constant. In $\textrm{Cd}_3\textrm{As}_2$, we found that the frequency of the LOB decreases monotonically with decreasing $\sigma$, approaching roughly 0.1 THz with $\sigma=0.01$ eV ($T_{\textrm{el}}=115$ K). In comparison, the well-known high-performance thermoelectric materials with soft optical phonons, such as $\textrm{Bi}_2\textrm{Te}_3$\cite{hellman2014phonon} and PbTe\cite{delaire2011giant}, have their LOBs around 1 THz. 

This strong dependence of optical phonon frequencies on $\sigma$ was also observed in graphene, where optical phonons at the $\Gamma$ and K points in the Brillouin zone showed similar behaviors\cite{piscanec2004kohn}. In graphene, this phenomenon was attributed to the Kohn anomaly\cite{kohn1959image}. In 1959, Kohn pointed out that, in a metal, the resonance between lattice vibrations and the Fermi surface can lead to nonanalytic points in the phonon dispersion. These anomalies should happen when $\mathbf{q}=2\mathbf{k}_\textrm{f}$, where $\mathbf{q}$ is the phonon wavevector and $\mathbf{k}_\textrm{f}$ is the Fermi wave vector\cite{kohn1959image}. Here, $\mathbf{q}=2\mathbf{k}_\textrm{f}$ is the maximum wave vector of a phonon that can be involved in a scattering event with two electronic states on the Fermi surface. In graphene, the Fermi surface is virtually two Dirac points located at K and $\textrm{K}'$, therefore Kohn anomaly can only happen at $\Gamma$ (intra-node scattering) and K points (inter-node scattering between K and $\textrm{K}'$). In a Dirac material, the size of the Fermi surface is highly sensitive to the electronic smearing parameter $\sigma$, which explains the strong dependence of the optical phonon frequency in graphene on $\sigma$. In $\textrm{Cd}_3\textrm{As}_2$, the two Dirac nodes are located at $(0,0,\pm 0.043\  \textrm{\AA}^{-1})$. Therefore, Kohn anomalies should potentially appear at $\Gamma$ (intra-node scattering) and $\mathbf{q}_0=(0,0,\pm 0.086\  \textrm{\AA}^{-1})$ (inter-node scattering), as marked in Fig. \ref{fig1}(c) and \ref{fig2}(b). In addition to the drastic phonon softening at $\Gamma$, a slight dip of one optical phonon branch at $(0,0,\pm 0.086\  \textrm{\AA}^{-1})$ is observed, again signaling the presence of Kohn anomaly. We emphasize here that the current calculation can only capture the Kohn anomaly due to static electronic screening of lattice vibrations, while a full treatment including dynamic screening effect will require the calculation of dynamic electron-phonon coupling that was done in the case of graphene\cite{lazzeri2006nonadiabatic} but is currently inaccessible for $\textrm{Cd}_3\textrm{As}_2$ given its complex structure.         

To verify the existence of the group of low-frequency optical phonons at $\Gamma$, we conducted temperature-dependent Raman measurement of high-quality $\textrm{Cd}_3\textrm{As}_2$ thin films grown by MBE\cite{schumann2016molecular}. The $\textrm{Cd}_3\textrm{As}_2$ layer is $\sim$30 nm thick and grown on a (111) GaSb/GaAs substrate. Growth details and electrical characterization have been reported elsewhere\cite{schumann2016molecular}. We used unpolarized excitation with 488 nm wavelength and 65 mW power for the Raman measurement at 300 K, 120 K and 77 K. To rule out the influence of the substrate, we also measured the Raman spectrum of the substrate with the same configuration (provided in Supplementary Information). As compared to existing reports of Raman spectra of bulk $\textrm{Cd}_3\textrm{As}_2$\cite{jandl1984raman,hosseini2016large,sharafeev2017optical,weszka1986some}, our measurement, as shown in Fig. 3, revealed the existence of a low-frequency optical phonon mode near 20 $\textrm{cm}^{-1}$ at all three temperatures and another mode near 15 $\textrm{cm}^{-1}$ at 120 K and 77 K. The spectral region with smaller Raman shift is masked by a strong background signal due to quasielastic electronic scatterings\cite{sharafeev2017optical} and thus the LOB cannot be resolved. The frequencies of both resolved modes agree well with our first-principles calculation. In addition, we observed a decreasing trend of the frequency of the phonon mode near 20 $\textrm{cm}^{-1}$ as the temperature was lowered, which was in contrast to the opposite temperature dependence of optical phonons with higher frequencies\cite{sharafeev2017optical} caused by thermal expansion.   
\begin{figure}[h]
\centering
\graphicspath{{./}}
\includegraphics[width=\columnwidth]{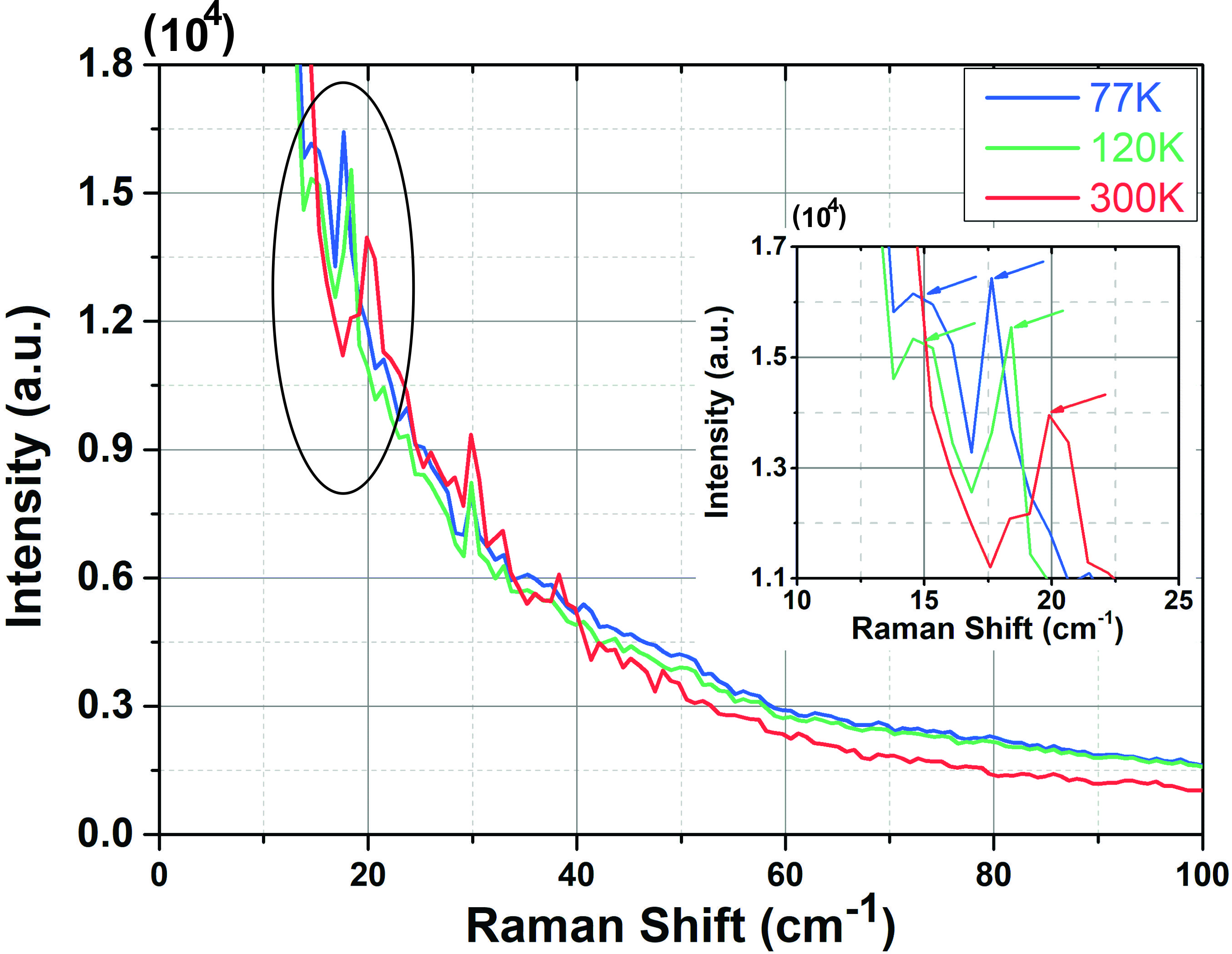}
\caption{The Raman spectrum of $\textrm{Cd}_3\textrm{As}_2$ measured at three different temperatures. The inset shows the enlarged low-frequency region and the arrows mark the Raman peaks resolved by the measurement.}
\label{fig3}
\end{figure}

\begin{figure*}[htb]
\centering
\graphicspath{{./}}
\includegraphics[width=\textwidth]{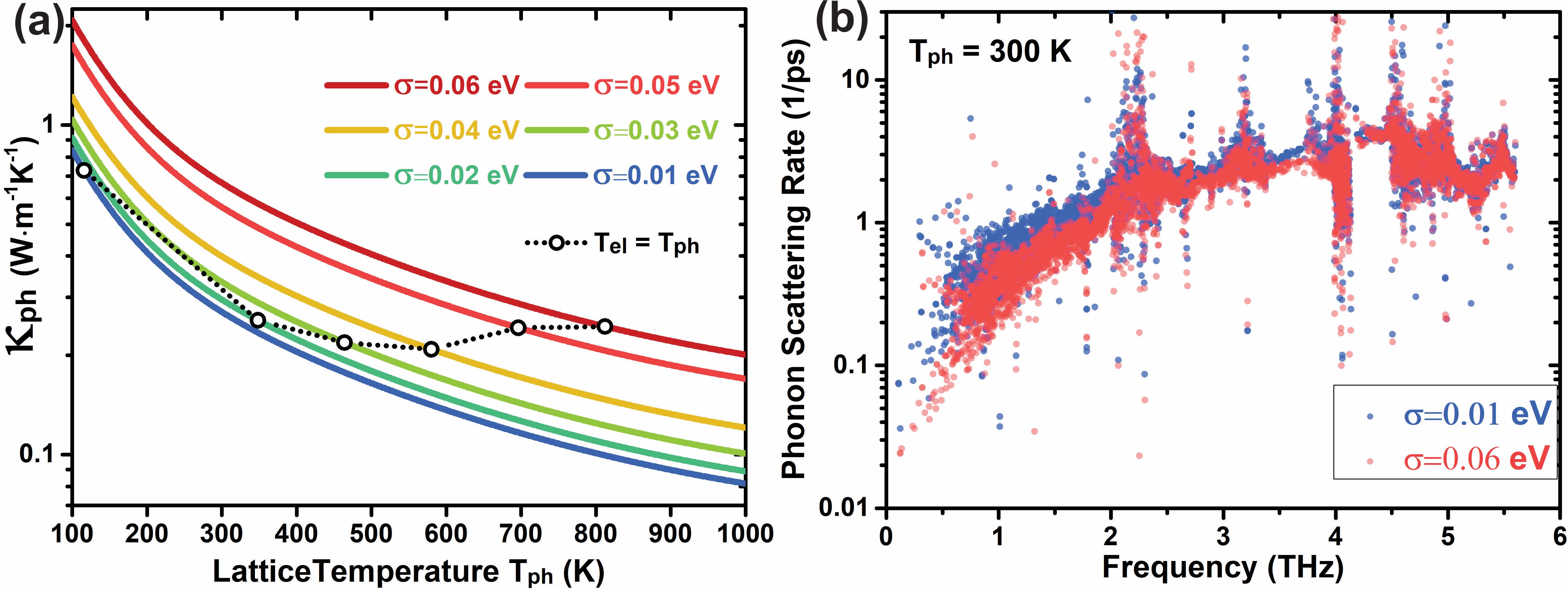}
\caption{(a) The calculated lattice thermal conductivity of $\textrm{Cd}_3\textrm{As}_2$ as a function of lattice temperature based on the phonon dispersion relations calculated using different $\sigma$ values. The open circles and dotted line mark the lattice thermal conductivity values obtained when $T_{\textrm{el}}=T_{\textrm{ph}}$. (b) The phonon-phonon scattering rates as a function of phonon frequency at $T_{\textrm{ph}}=300$ K with two different $\sigma$ values}.
\label{fig4}
\end{figure*}  

Finally, we investigated how the existence of soft optical phonon modes affects the lattice thermal conductivity of $\textrm{Cd}_3\textrm{As}_2$. It is well understood that low-frequency optical phonon modes can significantly increase the available phase space
for phonon-phonon scattering of heat-carrying acoustic phonons, which is responsible for the low lattice thermal conductivity of the best thermoelectric materials, such as $\textrm{Bi}_2 \textrm{Te}_3$\cite{hellman2014phonon,lee2014resonant,yue2018ultralow}, PbTe\cite{delaire2011giant,li2014phonon} and SnSe\cite{li2015orbitally}. In these materials, the soft optical phonon mode usually signals their proximity to a structural phase transition. Here, we expect the low-lying optical phonon modes in $\textrm{Cd}_3\textrm{As}_2$ will have a similar effect. Due to the complexity and low symmetry of the primitive cell of $\textrm{Cd}_3\textrm{As}_2$, it is computationally intractable to calculate the anharmonic force constants and phonon-phonon scattering rates using the full cell. To overcome this obstacle, we examined the thermal transport properties of the crystal constructed from the 10-atom substructures. The similarity between the dispersion relations of the long-wavelength phonons calculated using the full cell and the 10-atom substructure, as observed in Fig. \ref{fig2}(a), provides justification to this approach, since these long-wavelength phonons are major heat carriers in $\textrm{Cd}_3\textrm{As}_2$. In particular, we expect this approach can qualitatively capture the physics of the soft phonon modes and their impact on thermal transport. Another unique aspect is the strong dependence of the LOB frequency on the electronic smearing parameter $\sigma$, which implies that the actual phonon dispersion of $\textrm{Cd}_3\textrm{As}_2$ will be highly sensitive to temperature. To include this effect, we analyzed thermal transport properties based on phonon dispersion relations assuming different values of $\sigma$. We calculated the lattice thermal conductivity $\kappa_{\textrm{ph}}$ by solving phonon Boltzmann transport equation (BTE) iteratively using ShengBTE\cite{ShengBTE}. The anharmonic third-order IFCs were calculated using the finite displacement method\citep{bte,3RD-IFCs}. A $2\times 2\times 2$ supercell was used for the calculation and the interactions between atoms were taken into account up to sixth nearest neighbors. The convergence of $\kappa_{\textrm{ph}}$ with the interaction distance between atoms was checked. The q-space (phonon momentum space) sampling grid was set to $\rm 12 \times 12 \times 12$. The grid density convergence of all cases were examined and all the $\kappa_{\textrm{ph}}$ reported here are converged values.  

Our results are shown in Fig. \ref{fig4}(a). Here, the lattice temperature $T_{\textrm{ph}}$ determines the Bose-Einstein distribution of the phonons involved in the phonon-phonon scattering calculation while $\sigma$ dictates the broadening of the Fermi-Dirac distribution of electrons used in the DFT calculation. For a given $\sigma$, the lattice thermal conductivity decreases with an increasing lattice temperature, which is typical for crystalline materials limited by phonon-phonon Umklapp scattering. At 300 K, our calculated lattice thermal conductivity is in the range of 0.3 to 0.9 W/mK, in good agreement with experimental reports\cite{spitzer1966anomalous,armitage1969thermal,wang2018magnetic}. For a given lattice temperature, the calculated lattice thermal conductivity is lower with a smaller $\sigma$. This is expected as a smaller $\sigma$ leads to a lower LOB frequency, which in turn causes stronger scattering of low-frequency acoustic phonons. We confirm this hypothesis by calculating the phonon-phonon scattering rates at a fixed lattice temperature but with two different $\sigma$ values, as shown in Fig. \ref{fig4}(b). While the scattering rates of higher-frequency optical phonons above 2 THz are almost identical in the two cases, the scattering rates of low-frequency heat-carrying acoustic phonons are significantly enhanced with a smaller $\sigma$. 

In order to make a direct connection to experimental results, both the phonon-phonon Umklapp scattering and the $\sigma$-dependent optical phonon frequency need to be taken into account. The counteraction of the two effects is expected to generate a non-monotonic temperature dependence of the lattice thermal conductivity in $\textrm{Cd}_3\textrm{As}_2$. A rigorous treatment would entail a dynamic calculation\cite{lazzeri2006nonadiabatic,giustino2017electron} that is beyond the current scope. However, to qualitatively capture the interplay of the two effects, we can compare the lattice thermal conductivity values calculated when the fictitious electronic temperature $T_\textrm{el}$ determined by $\sigma$ is set equal to the lattice temperature $T_\textrm{ph}$. These values are marked in Fig. \ref{fig4}(a) with open circles and connected by a dotted line. Below 450 K, the temperature dependence of phonon-phonon Umklapp scattering is more prominent, and thus the lattice thermal conductivity decreases with temperature. Above 450 K, however, the increasing optical phonon frequency plays a more important role and the lattice thermal conductivity starts to increase with temperature. This physical picture provides a qualitative explanation for the increasing trend of the lattice thermal conductivity of $\textrm{Cd}_3\textrm{As}_2$ with temperature as observed experimentally\cite{wang2018magnetic}, while the quantitative difference from the experimental result might be due to the simplified treatment used here.

In summary, we investigated the lattice dynamics and thermal transport of Dirac semimetal $\textrm{Cd}_3\textrm{As}_2$ using first-principles calculation and Raman measurement. We identified the existence of soft optical phonons likely due to Kohn anomaly associated with the Dirac nodes. Based on the observation of the soft modes, we explained the ultralow lattice thermal conductivity of $\textrm{Cd}_3\textrm{As}_2$ due to soft-mode-enhanced phonon-phonon scatterings. We further suggested that the interplay of phonon-phonon Umklapp scattering and the optical phonon frequency can potentially explain the anomalous temperature dependence of the lattice thermal conductivity of $\textrm{Cd}_3\textrm{As}_2$ as observed experimentally. Our work exemplifies the rich phonon physics in topological materials.

\begin{acknowledgments}
This work is based on research supported by the U.S. Department of Energy, Office of Basic Energy Sciences, Division of Materials Science and Engineering through the Early Career Research Program under the award number DE-SC0019244. B.L. acknowledges the support provided by the Regents' Junior Faculty Fellowship from the University of California, Santa Barbara (UCSB). M.G., T.S., and S.S. acknowledge support through a Vannevar Bush Faculty Fellowship program by the U.S. Department of Defense (Grant number N00014-16-1-2814).
\end{acknowledgments}

\appendix


\bibliography{references.bib}

\end{document}